\documentclass[aps,prl,twocolumn,superscriptaddress,noshowpacs]{revtex4-2}
\usepackage[utf8]{inputenc}
\usepackage[T1]{fontenc}
\usepackage{graphicx}
\usepackage{xcolor}
\usepackage{physics}
\usepackage{bm}
\usepackage{amsfonts}
\usepackage{hyperref}
\newcommand{\tcr}[1]{\textcolor{black}{#1}}

\begin{document}

\title{Soliton formation in a bound state in the continuum GaN waveguide polariton laser}

\author{V. Develay}
\altaffiliation{These authors have contributed equally to the present work.}
\affiliation{Laboratoire Charles Coulomb, Université de Montpellier, CNRS, 34095 Montpellier, France}

\author{O. Bahrova}
\altaffiliation{These authors have contributed equally to the present work.}
\affiliation{Institut Pascal, PHOTON-N2, Universit\'e Clermont Auvergne, CNRS, Clermont INP,  F-63000 Clermont-Ferrand, France}
\affiliation{B. Verkin Institute for Low Temperature Physics and
Engineering of the National Academy of Sciences of Ukraine, 47
Nauky Ave., Kharkiv 61103, Ukraine}

\author{I. Septembre}
\affiliation{Naturwissenschaftlich-Technische Fakultät, Universität Siegen, Walter-Flex-Straße 3, 57068 Siegen, Germany}

\author{D. Bobylev}
\affiliation{Institut Pascal, PHOTON-N2, Universit\'e Clermont Auvergne, CNRS, Clermont INP,  F-63000 Clermont-Ferrand, France}

\author{R. Baye}
\affiliation{Laboratoire Charles Coulomb, Université de Montpellier, CNRS, 34095 Montpellier, France}

\author{C. Brimont}
\affiliation{Laboratoire Charles Coulomb, Université de Montpellier, CNRS, 34095 Montpellier, France}

\author{L. Doyennette}
\affiliation{Laboratoire Charles Coulomb, Université de Montpellier, CNRS, 34095 Montpellier, France}

\author{B. Alloing}
\affiliation{CRHEA, Universit\'e Cote d'Azur, CNRS, Rue Bernard Gregory, 06560 Valbonne, France}

\author{H.~Souissi}
\affiliation{Centre de Nanosciences et de Nanotechnologies, CNRS, Universit\'e Paris-Saclay, France}

\author{E.~Cambril}
\affiliation{Centre de Nanosciences et de Nanotechnologies, CNRS, Universit\'e Paris-Saclay, France}

\author{S.~Bouchoule}
\affiliation{Centre de Nanosciences et de Nanotechnologies, CNRS, Universit\'e Paris-Saclay, France}

\author{J.~Z\'u\~niga-P\'erez}
\affiliation{CRHEA, Universit\'e Cote d'Azur, CNRS, Rue Bernard Gregory, 06560 Valbonne, France}
\affiliation{MajuLab, International Research Laboratory IRL 3654, CNRS, Université Côte d’Azur, Sorbonne Université, National University of Singapore, Nanyang Technological University, Singapore, Singapore}

\author{D. Solnyshkov}
\affiliation{Institut Pascal, PHOTON-N2, Universit\'e Clermont Auvergne, CNRS, Clermont INP,  F-63000 Clermont-Ferrand, France}
\affiliation{Institut Universitaire de France (IUF), 75231 Paris, France}

\author{G. Malpuech}
\email{guillaume.malpuech@uca.fr}
\affiliation{Institut Pascal, PHOTON-N2, Universit\'e Clermont Auvergne, CNRS, Clermont INP,  F-63000 Clermont-Ferrand, France}

\author{T. Guillet}
\email{Thierry.Guillet@umontpellier.fr}
\affiliation{Laboratoire Charles Coulomb, Université de Montpellier, CNRS, 34095 Montpellier, France}

\begin{abstract}
We study polaritonic bound states in the continuum (BIC) created in GaN waveguides. The existence of symmetry-protected BICs is confirmed by the suppression of light emission and the observation of a polarization vortex in momentum space. Upon increasing the pumping, polariton population accumulates at the BIC and we observe polariton lasing from the blueshifted BIC states. The assessment of the polariton BIC emission energy \textcolor{black}{and of its real and momentum space wavefunctions} as a function of pumping power, i.e. of polariton density, indicates the formation of a bright soliton above the lasing threshold. Soliton formation at the BIC is induced by the combination of negative mass BIC and of repulsive polariton-polariton \textcolor{black}{and polariton-reservoir} interactions.
\end{abstract}
\maketitle

Bound states in the continuum (BICs) have become a particularly active topic of research in the last years~\cite{hsu2016bound}. Various states with suppressed radiative emission were known and studied in optics for a long time. However, only their recent interpretation as BICs, known from quantum mechanics~\cite{neumann1929merkwurdige,friedrich1985interfering}, and their description in terms of symmetry~\cite{Sadrieva2019,Overvig2020} and topology~\cite{zhen2014topological} has allowed to engineer and exploit them in numerous applications. These cover communications~\cite{dreisow2009adiabatic,gentry2014dark}, lasers~\cite{rybin2017supercavity,kodigala2017lasing,ha2018directional}, filters~\cite{foley2015normal} and sensors~\cite{foley2014symmetry}. Topologically-protected BICs have been demonstrated in many configurations~\cite{doeleman2018experimental,yoda2020generation,shen2024topologically}, with their topology being often linked to the polarization degree of freedom, which is at the origin of many  phenomena in photonics~\cite{lu2014topological,solnyshkov2021microcavity} thanks to the ubiquitous transverse-electric/transverse-magnetic (TE-TM) spin-orbit coupling~\cite{bliokh2008geometrodynamics,bliokh2015spin,gianfrate2020measurement}. In the simplest case, the states around the BIC exhibit a so-called polarization vortex (e.g. TE-polarized), which makes impossible to define the polarization of the BIC itself, which appears as a singularity~\cite{zhen2014topological}. From the practical point of view~\cite{kang2023applications}, BICs are often used to ensure lasing at a well-controlled energy and to reduce its nonlinear threshold~\cite{kodigala2017lasing,yu2021ultra,chen2022observation}. 

In parallel to these developments in nanophotonics, strong light-matter coupling~\cite{hopfield1958theory} was studied as a means of achieving large nonlinearities~\cite{baas2004optical,emmanuele2020highly} in optical systems thanks to the interactions of matter-dressed photons, called polaritons~\cite{basov2020polariton}. In particular, exciton-polaritons have demonstrated a particularly rich spectrum of possibilities, from the observation of Bose-Einstein condensation~\cite{kasprzak2006bose} and superfluidity~\cite{amo2009superfluidity} to topological phenomena, as optical spin Hall effect~\cite{kavokin2005optical,leyder2007observation} and anomalous Hall effect~\cite{gianfrate2020measurement}. Note that while these effects were initially observed at cryogenic temperatures, they have been extended up to room temperature~\cite{christopoulos_room-temperature_2007,lerario2017room,liang2024polariton,shi2025coherent}. Interestingly, patterning of photonic structures displaying simultaneously light-matter coupling and spin-orbit coupling is nowadays well-mastered in different material platforms~\cite{jacqmin2014direct,st2017lasing,klembt2018exciton,su2020observation}, enabling a controlled combination of the two aspects.

In the field of exciton-polaritons, BICs have been considered as means to control the emission rate of polariton structures for at least a decade~\cite{solnyshkov2011polariton}, and a number of results have been achieved since then: polariton interactions in BIC states have been measured~\cite{kravtsov2020nonlinear}, a topological polarization vortex has been observed~\cite{dang2022realization} and polariton condensation on BIC states has been experimentally demonstrated recently, first at cryogenic~\cite{ardizzone2022polariton} and then at room~\cite{berghuis2023room,wu2024exciton,yan2025topologically} temperature. Even combined states of exciton-polaritons and plasmonic BICs have been studied~\cite{luo2025room}. Recently, polariton condensates in BICs were used to claim supersolidity~\cite{trypogeorgos2025emerging}. In this context, the general effects of polariton-polariton (and other more general) interactions have been studied for BICs, both theoretically and experimentally~\cite{yang2001embedded,Bulgakov2010,Krasikov2018,dolinina2020bic,hu2021nonlinear,dolinina2021interactions,riminucci2023bose,grudinina2023collective,gianfrate2024reconfigurable,berghuis2024condensation,Padhy2026}. Predicted nonlinear effects include mechanisms of the BIC formation (embedded solitons~\cite{yang2001embedded}), bistability~\cite{Bulgakov2010} and instabilities~\cite{Krasikov2018}, as well as creation of various solitonic states~\cite{dolinina2020bic}. In particular, the formation of a polariton condensate soliton on a BIC state has been predicted~\cite{septembre2024soliton}, with its change of shape, angular emission pattern and blue shift being influenced by the interactions. However, an experimental demonstration of the soliton formation remains elusive.

In this work, we report the first experimental observation of polariton condensation on a BIC in a GaN patterned structure and we demonstrate its experimental fingerprints, including the suppression of far-field emission and the appearance of a polarization vortex, which ensures the topological protection of the BIC. \textcolor{black}{Most importantly, we report k-space and 2-dimensional (2D) real space images of the BIC condensate. They show gap soliton states having a lateral size of about 3~$\mu$m. The localization mechanisms are the attractive exciton reservoir and polariton-polariton interaction both acting on negative mass particles.}
The soliton formation is particularly important because its presence affects the angular emission diagram of the BIC condensate and modifies the light extraction efficiency.

\begin{figure}
    \centering
    \includegraphics[width=1.0\linewidth]{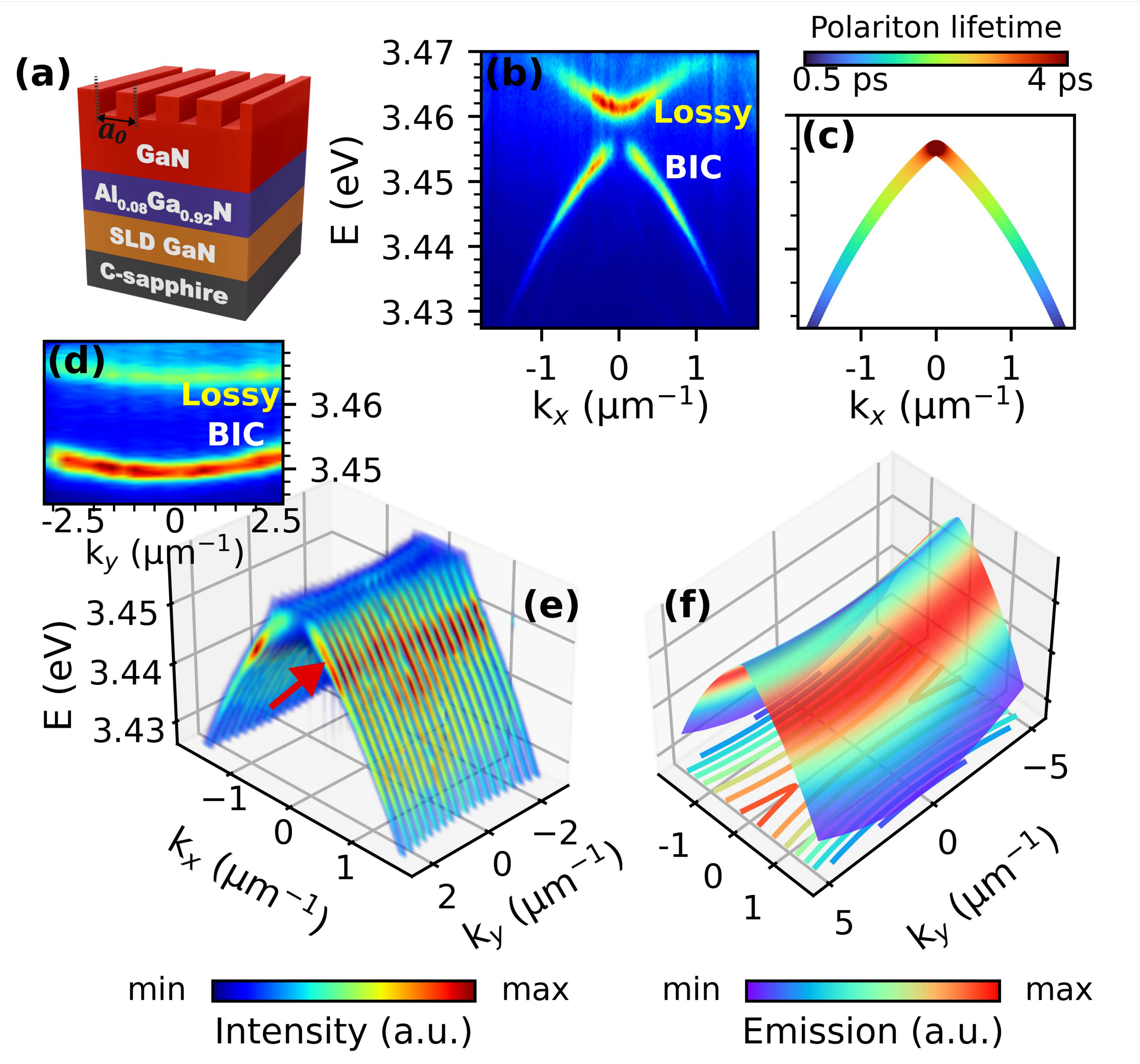}
    \caption{Bound state in continuum in a GaN polariton waveguide. (a) Scheme of the experimental structure. (b,c) Dispersion relation of the polaritonic guided modes measured along the in-plane direction perpendicular to the grating and exhibiting a BIC (b - experiment, c - theory (taking exciton lifetime 2 ps). (d) Dispersion relation of the polaritonic guided modes along the transverse direction, i.e. parallel to the grating, for a nonzero  $k_x$, $k_x=0.5\mu m^{-1}$, showing parabolic behaviour; (e,f) 3D dispersion relations, $E(k_x,k_y)$ (e - experiment, f - theory) around $k_x=k_y=0$. The red arrow indicates the wave vector value, which corresponds to the panel (d). }
    \label{fig1}
\end{figure}

\textit{Polariton BIC.} We study a waveguide of a wide-bandgap semiconductor (GaN) exhibiting pronounced excitonic resonances which are able to strongly couple with photonic modes even at room temperature~\cite{semond2005strong,brimont2020strong}. The scheme of the sample is shown in Fig.~\ref{fig1}(a). The sample contains a patterned 232~$nm$-thick GaN waveguide layer grown by metalorganic vapour phase epitaxy on top of an AlGaN bottom cladding, itself grown on a GaN-on-Sapphire template displaying a low dislocation density of the order of $3\times10^8$cm$^{-2}$. The 1D lattice is fabricated by electron beam lithography and inductive plasma etching~\cite{bouchoule2018technologies,souissi_Ridge_2022}, with the parameters of the patterning (period $a_0=132 \ nm$, filling factor 60\%, etching depth 55~$nm$) chosen in order to obtain a BIC for photonic modes at an energy allowing their strong coupling with the GaN exciton resonances~\cite{septembre2024soliton}.

Optical experiments were performed at 70~K using quasi-cw excitation at 349~nm, slightly above the energy of the exciton reservoir, with a Gaussian pump spot of about 28~$\mu m$ FWHM, much smaller than the 200~$\mu m-$wide square patterned area defining the BIC.
The photoluminescence (PL) is collected by a 10x microscope objective in a Fourier imaging scheme, giving access to the intensity as a function of energy and wave vector.

Fig.~\ref{fig1} presents an overview of the polaritonic waveguide  employed and its full dispersion in the linear regime. Along the modulation direction ($k_x$), as shown in Fig.~\ref{fig1}(b), the BIC appears as a dark spot in the lower branch because the far-field emission from the BIC is topologically suppressed (see below). The experimental results are in agreement with numerical simulations of the  photonic structure performed with COMSOL Multiphysics (2D eigenvalue problem with periodic Floquet boundary conditions,  see \cite{suppl} for details). In the absence of excitons, the simulation shows the standard divergence of radiative lifetime for the BIC state. A more realistic simulation, including finite size and coupling to excitons, is presented in Fig.~\ref{fig1}(c). 
The BIC appears as a dark red spot in the lower band, exhibiting the longest lifetime among all states of this band. We stress that the topological suppression applies only to radiative losses and the associated far field emission, while other losses are not suppressed, which is why the lifetime in realistic simulations is actually finite. Moreover, in a finite-size system even radiative losses are not suppressed completely. Nevertheless, from the experimental point of view, the emission from the BIC is much weaker than the background signal, so it can be considered as totally suppressed. 

Along the perpendicular in-plane direction, the photonic and polaritonic modes exhibit a positive effective mass. Fig.~\ref{fig1}(d) presents the PL intensity as a function of energy $E$ and transverse wave vector $k_y$ for a non-zero $k_x=0.5$~$\mu$m$^{-1}$, allowing thereby to observe a non-zero signal from the BIC branch. Overall, the BIC is located at a saddle point of the polaritonic dispersion, Fig.~\ref{fig1}(e), corresponding to a maximum along the  $k_x$ direction and a minimum along the $k_y$ direction in momentum space. Such a saddle-point configuration is completely reproduced by the numerical simulations, Figure~\ref{fig1}(f), where the emission is calculated taking into account the emission rate $\Gamma(k_x,k_y)$ and the polariton population obtained by resonant scattering from the excitonic reservoir with LO phonons~\cite{Levrat2010,jamadi2016polariton,septembre2024soliton}.

\begin{figure}
    \centering
    \includegraphics[width=1.0\linewidth]{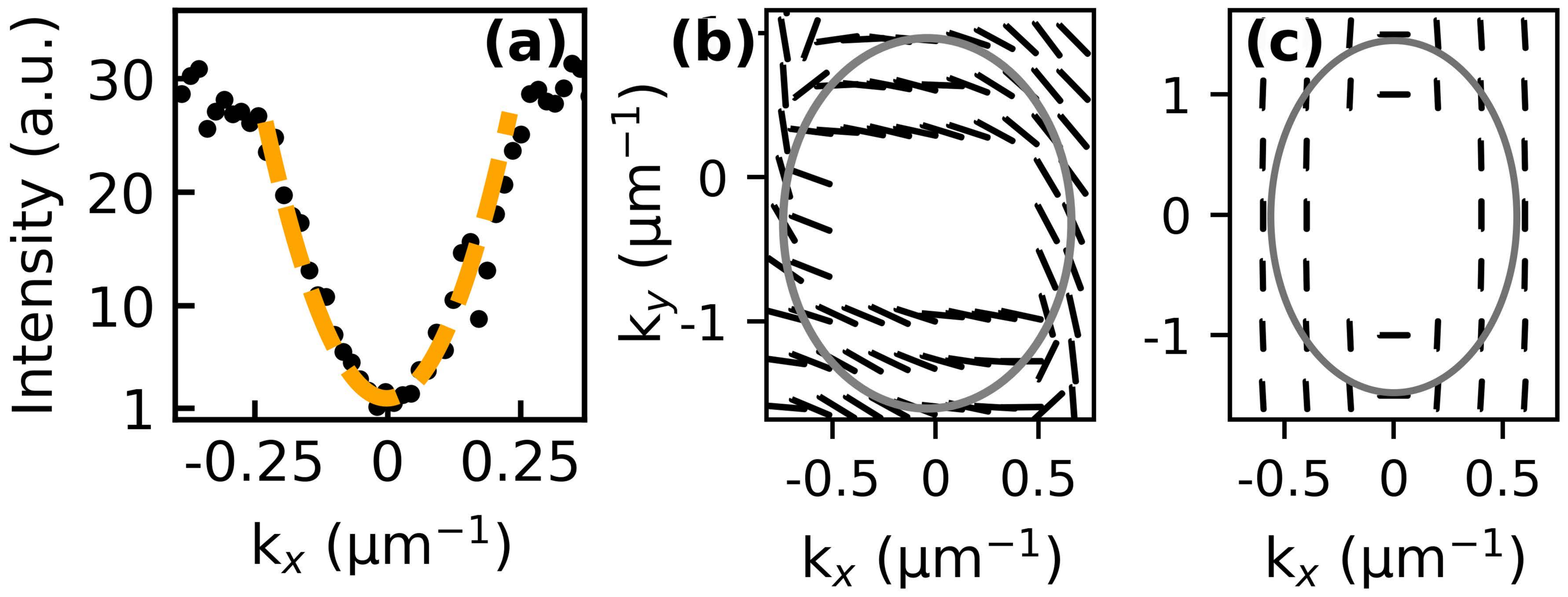}
    \caption{The BIC properties. (a) Intensity of emission as a function of wave vector and its quadratic fit; (b,c) Linear polarization orientation (b - experiment, c - theory)
    demonstrating a vortex. In experiment, the signal from the central part is too weak. The elliptic curve is a guide for the eyes allowing to follow the tangential orientation of the linear polarization.}
    \label{fig2}
\end{figure}

The key property of the BIC is its quenched emission rate at $k_x=0$. This is demonstrated in Fig.~\ref{fig2}(a), showing a drastic change (around 30 times) of the PL intensity near the BIC. The intensity of the PL emission along the BIC band is plotted as a function of the wave vector $k_x$ (for $k_y=0$, which corresponds to the minimal value of radiative losses associated to $k_x=0$, as shown in Fig.\ref{fig1}(e,f)), overlaid with a parabolic fit. This dependence follows the theoretically expected behavior. However, for large  $k_x$ values this ideal dependence is perturbed by two distinct effects: first, at large wavevectors the BIC band couples with bands of different polarization and, second, for these wavevectors the population of the polaritonic states is largely reduced due to the increase of the energy detuning with respect to the optimum LO-phonon resonance, which is the most efficient relaxation channel for polaritons in GaN, as previously demonstrated \cite{jamadi2016polariton}.

Another important feature of non-accidental BICs is their topological protection, which guarantees the existence of a BIC. This is an important design advantage when the possible imperfections of the sample are taken into account. Fig.~\ref{fig2}(b,c) demonstrates the polarization vortex (experiment and theory, respectively) formed around the BIC, which ensures its protection. In these panels we plot the orientation of the linear polarization extracted from the eigenstates along the BIC band. To do so, we fit the polarization-resolved spectra at each wave vector and construct the components of the Stokes vector (see Supplementary for more details~\cite{suppl}). We stress that because of a low PL intensity emission very near the BIC, in experiments we need to restrict the measurement to k-states with sufficiently large intensity, Fig.~\ref{fig2}(b).
Together, the two experimental observations (suppressed emission and polarization vortex) confirm the BIC nature of the polaritonic mode under study.

\begin{figure}
    \centering
    \includegraphics[width=1.0\linewidth]{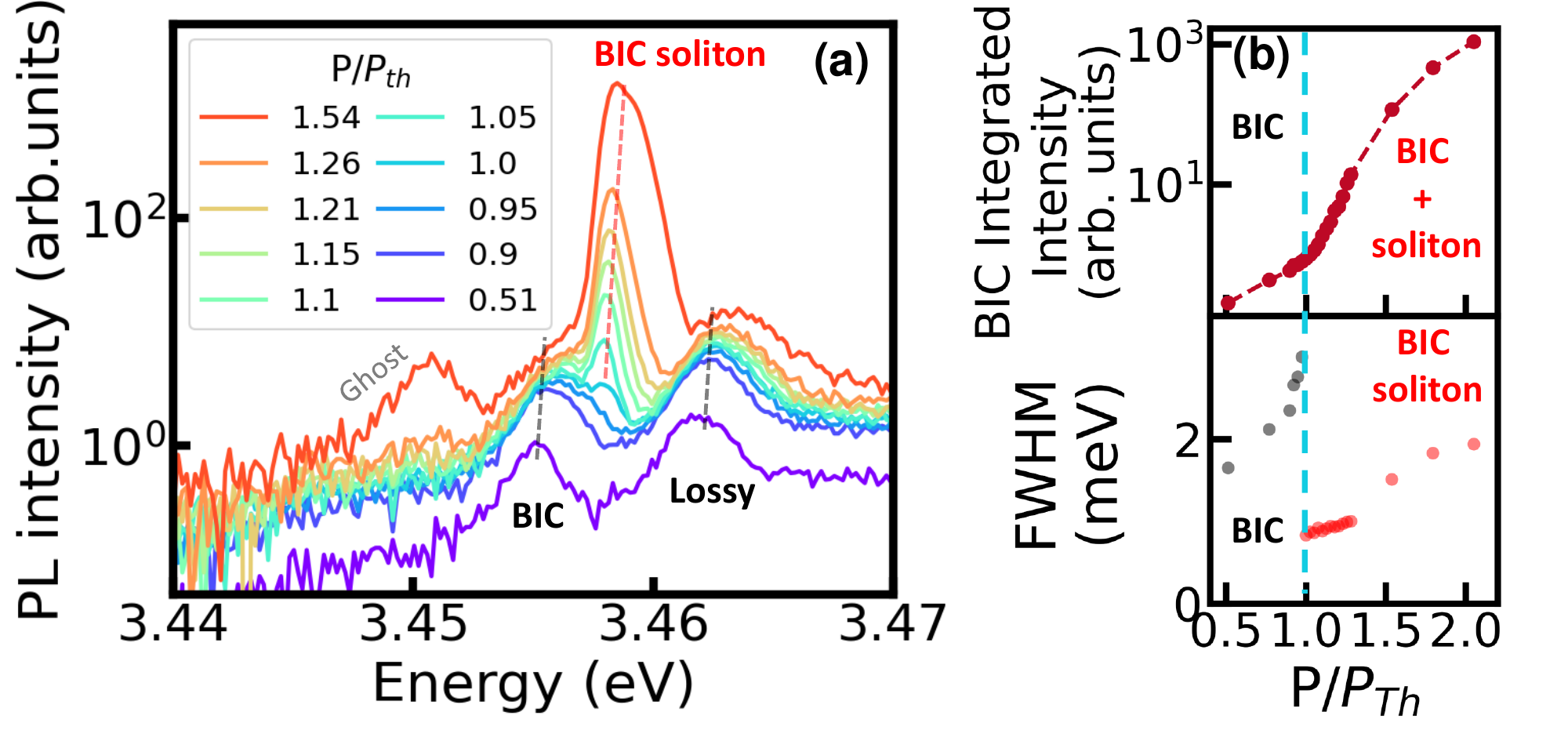}
    \caption{BIC lasing. (a) Experimental spectra at $k_x=$0.15 $\mu$m$^{-1}$ for different pumping powers relative to the lasing threshold, demonstrating the build-up of the condensate emission from the BIC \textcolor{black}{(28~$\mu$m pump)}. At high power, a ghost is detected at 3.45 eV because of the intense BIC peak; \textcolor{black}{(b) [Top] BIC emission intensity as a function of the pumping power between 3.456 and 3.462~eV, exhibiting a typical threshold curve; [Bottom] Linewidth of the BIC peak (light black) and BIC soliton peak (light red), exhibiting a reduction at the lasing threshold.}}
    \label{fig3}
\end{figure}

\textit{Polariton condensation on a BIC.} We now analyze the emission of our structure with the increase of the pumping power.
The evidence for polariton condensation on the BIC state is shown in Fig.~\ref{fig3}. Panel (a) shows several PL spectra for different pumping powers expressed as multiples of the threshold power $P_{th}$. The latter is defined in Fig.~\ref{fig3}(b), which shows the integrated PL intensity as a function of pumping power. The output-input curve exhibits a typical change of slope corresponding to the onset of stimulated scattering towards the polariton BIC state (polariton condensation) accompanied by coherent emission (lasing).
The BIC emission is observed due \tcr{to} the finite k-space broadening of the polariton condensate wavefunction, broadening which is moreover enhanced by the formation of interaction-induced gap soliton, as we show below.
The superlinear increase of the PL at the BIC energy together with its blue shift due to polariton-polariton interactions are also visible in Fig.~\ref{fig3}(a), and will be analyzed in detail in the following section. 

\begin{figure}
    \centering
    \includegraphics[width=1.0\linewidth]{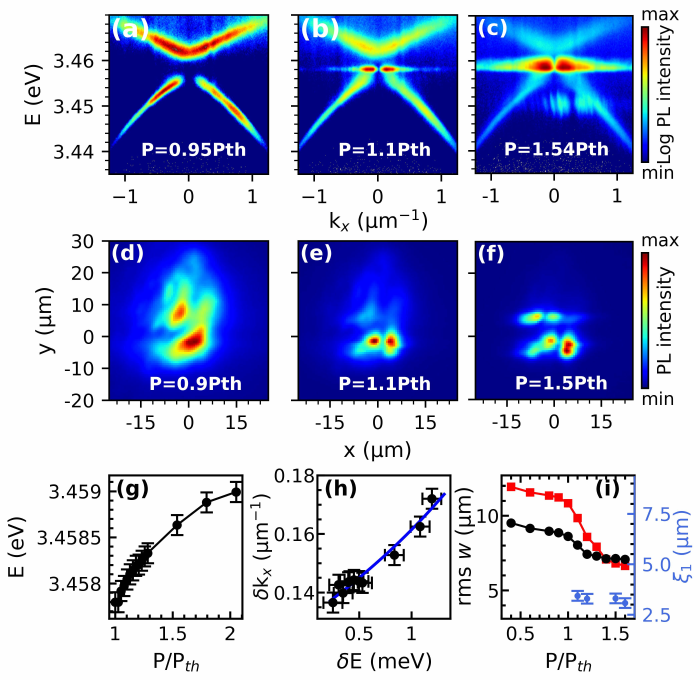}
    \caption{BIC solitons. Experimental (a,b,c) \textcolor{black}{dispersions along the in-plane direction demonstrating the condensation at a BIC state and illustrating the blueshift of the BIC above threshold and its broadening in k-space due to the interactions (log-scale). \tcr{(d,e,f)} Experimental 2D real space emission intensities.\tcr{(g)}  Energy of the BIC as a function of normalized pumping power. (h) Size of the soliton in k-space as a function of the soliton blueshift energy (dots - experiment, blue curve - theory). (i) Root mean square width calculated along $x$ (black points) and $y$ (red squares). Blue dots correspond to the real-space soliton size (along the BIC direction) obtained from the fit.}}
    \label{fig4}
\end{figure}

\begin{figure}
    \centering
    \includegraphics[width=1.0\linewidth]{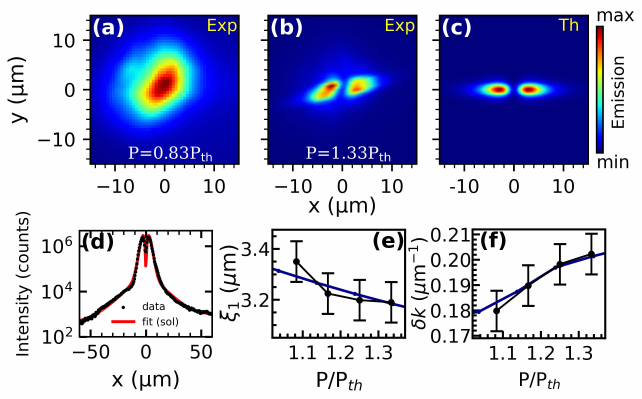}
    \caption{\textcolor{black}{Small pumping spot. Experimental (a-b) and theoretical (c) 2D real space emission intensities. Extracted from the fit (data (black dots) together with the fit (red curve) is presented in (d)) real space width of the gap soliton (black dots with errorbars) versus pump is plotted in (e) (blue curve -- theory) and the corresponding extracted from the fit momentum space widths are in (f).  }}
    \label{fig5}
\end{figure}

\textit{BIC soliton.} A BIC polariton Bose-Einstein Condensate (BEC) formed in a negative mass state as in the present case is expected to form a bright \textcolor{black}{gap} soliton, as theoretically described in \cite{septembre2024soliton}. 
\textcolor{black}{Indeed, the repulsive interactions between polaritons and with the reservoir become attractive for negative mass states, as initially demonstrated for cavity polaritons lattices \cite{tanese2013polariton}.
We performed experiments using pumping spots of 28~$\mu$m and 9~$\mu$m, shown in Fig.~\ref{fig4} and Fig.~\ref{fig5} respectively.}

Fig.~\ref{fig4}(a) shows experimental PL emission intensity around the BIC state below threshold (similar to Fig.~\ref{fig1}(b)). Figure~\ref{fig4}(b) shows the emission from the same states slightly above threshold, when the condensate becomes visible in the gap between the BIC and the lossy bands. Its energy is already blueshifted with respect to the BIC band observed below threshold, and continues to blueshift 1 meV more at 2 $P_{th}$, as shown in Fig.~\ref{fig4}(g). Note that the typical interaction energies observed so far in GaN waveguides are of the order of $\alpha n=0.5$~meV, where $\alpha$ is the effective polariton-polariton interaction constant mostly dominated by the saturation of the exciton oscillator strength and $n$ is their density~\cite{Ciers2020,souissi2024mode}, which is consistent with the blueshift here observed.

Figure~\ref{fig4}(h) shows the size of the emission in reciprocal space along $k_x$ extracted from the experimental PL images as a function of the energy (dots with error bars) and is well reproduced by the theory \cite{suppl}. \textcolor{black}{The \textit{k}-space broadening occurring above 1.5 $P_{th}$ suggests a real-space narrowing of the emission along the $x$-direction}.

\textcolor{black}{Fig.~\ref{fig4}(d-f) shows the real space emission (in a linear scale) measured at comparable powers and location with respect to (a-c) (but not simultaneously). Below threshold (Fig.~\ref{fig4}(d)), the intensity distribution appears quite inhomogeneous. The root mean square width (RMSw) of the total emission is 9 and 11.5~$\mu$m along $x$ and $y$, respectively (Fig.~\ref{fig4}(i)), which would correspond for a Gaussian distribution to full width at half maximum values of 21 and 27~$\mu$m, respectively.
The spatial distribution is similar for lower pumping. It is therefore not affected by the proximity of threshold or by complex behaviours of the excitonic reservoir \cite{perea2019exciton,bieker2015spatially}. It is certainly related to the sample inhomogeneity. Just above threshold (Fig.~\ref{fig4}(e)), the emission concentrates in a  state with 2 lobes, strongly localized both along $x$ and $y$ directions. The localization along $x$ is the result of the effective attraction of the reservoir potential and polariton-polariton interaction. This effective attraction is due to the negative mass of condensed polaritons. As shown in~\cite{suppl}, the emission with two lobes along $x$ corresponds to a wavefunction with zero nodes, filtered by the BIC band. The localization along $y$ is due to the blue shift combined with the finite lifetime and large mass along $k_y$ (\cite{suppl}). At higher pump power (Fig.~\ref{fig4}(f)), the emission pattern becomes more complex in real space. We can identify two main solitons at two different $y$. The description of soliton profiles along $x$ requires to consider both the BIC and the lossy band. The spatial distribution of emission along $x$, for a state localized either by self-interactions or by reservoir confinement only, reads~\cite{suppl}
\begin{equation}
I(x)\sim \beta I_{R,S}+(1-\beta)\exp(-2|x|/\xi_2)
\end{equation}
where 
\begin{equation}
I_{S}(x)\sim\tanh^2(x/\xi_1)\sech^2(x/\xi_1),\quad
I_R(x)\sim x^2e^{\frac{-x^2}{\sigma_E^2}}
\end{equation}
where $\beta$ is the contribution of the BIC band and $\xi_2$ the decay length associated with the presence of the second band. $\xi_1$ is the soliton healing length, which, in the conservative case and in the absence of reservoir, should read $\xi_1=\hbar/\sqrt{m\alpha n}$, where $n$ is the maximal particle density.
A fit of the intensity profile of the soliton located near $y=0$ using the wavefunction dominated by self-interactions is 1.3 times better (as measured by the rms error~\cite{suppl}) than using a fit based on the reservoir-confined wavefunction (for $1.5~P_{th}$). It 
yields a soliton width $\xi_1$ around 3.3$\pm$0.25~$\mu$m, which would correspond to a RMSw of about 4~$\mu$m (Fig.~\ref{fig4}(i), right and left axes, correspondingly). 
Within the pumping range 1.1-1.5~$P_{th}$, $\xi_1$ does not decrease significantly. The projection $1-\beta$ on the lossy band remains smaller than 0.1.}

\textcolor{black}{We next present measurements performed using a 9~$\mu$m spot, for which the transverse effects and the disorder play a smaller role. Fig.~\ref{fig5}(a-b) shows the real space emission at 0.9 and 1.3 $P_{th}$. The emission below threshold is quite homogeneous on this scale and a single soliton is visible above threshold.}
\textcolor{black}{The fit of the soliton profile at 1.3 $P_{th}$, using a hyperbolic wavefunction of interacting particles (Eqs.~(1,2)), is shown in Fig.~\ref{fig5}(d). The same fit based on a Gaussian wavefunction for non-interacting particles gives a rms error 1.6 times larger, confirming that the experimental profile is closer to a pure soliton than to a pure reservoir-localized state~\cite{suppl}.}
\textcolor{black}{Fig.~\ref{fig5}(e) shows the fitted value of $\xi_1$ versus pump power ($1-\beta$ remains smaller than 0.1), which shows a measurable narrowing, probably induced by the increase of polariton-polariton interaction. On the other hand, 
$\xi_1$ is comparable to the larger spot case (3.3~$\mu$m). This similarity is certainly due to the spatial fluctuations of the reservoir distribution visible on Fig~\ref{fig4}(d). The size of these fluctuations is comparable with the small spot size, yielding similar confinement potentials in both cases.
The width in k-space of the small spot case (extracted from the fit of the k-space emission) is shown in Fig.~\ref{fig5}(f). The product of the width in the two spaces is $2/\pi$, as expected for a soliton wavefunction. It proves that the emission in both spaces are directly related by a Fourier transform. For the large spot case, the k-space width was smaller than expected from the $\xi_1$-value. This is certainly due to the presence of several solitons along $y$ making the full real space emission effectively wider than $\xi_1$.}

\textcolor{black}{The modeling of such a soliton formation is performed through numerical simulations of polariton condensation in 2D with both the BIC and lossy bands in the presence of the polariton-polariton interactions and of the potential created by the excitonic reservoir using modified Gross-Pitaevskii equation with gain and losses ~\cite{suppl}.}
\textcolor{black}{These simulations allow us to compute the condensate wavefunction versus pumping, from which we can extract the emitted intensity comparable to experimental images (Fig.~\ref{fig5}(c)). The simulated widths in real and reciprocal space of the solitons versus pump are plotted in Fig.~\ref{fig5}(e,f) (blue line) and the agreement is good.
When the pumping power increases further, we find that the wavefunction projects more and more on the lossy band. The corresponding RMSw of the full wavefunction therefore passes by a minimum and then increases (see Fig.~S8 of \cite{suppl}).
The state becomes less favorable for condensation because it is losing its BIC character (i.e. infinite radiative lifetime) and it is less overlapping with the reservoir. Under these conditions, the condensation is expected to occur in the next soliton state of the potential. We indeed observed the onset of this phenomenon around 2~$P_{th}$, where a 3-lobe state (p-state modulated by the BIC) appears about 1.5~meV below the two-lobe state (see~Fig.~S11\cite{suppl})}

Overall, the long lifetime and localized nature of the BIC soliton in real space make it a promising candidate for photonic information treatment~\cite{peccianti2002all}, storage~\cite{sedov2014tunneling,pang2016all} and transfer~\cite{herr2016dissipative}. The angular broadening of polariton-based soliton BIC should be taken into account for laser applications and can turn to an advantage, increasing light extraction efficiency above threshold.

To conclude, we have fabricated a patterned GaN polariton waveguide with a polaritonic BIC state exhibiting topological protection. We have successfully observed polariton condensation on this state. Moreover, we have demonstrated one key feature of polariton condensates that differentiates them from bare-photon lasing: polariton-exciton and polariton-polariton interactions lead to the formation of a polariton BIC soliton and determine its energy and shape dependence on particle density, which  affects the angular and real-space distribution of the polariton laser emission. Our experimental results are in excellent agreement with the theory.

\begin{acknowledgments}
    
This work was supported by European Union's Horizon 2020 program, through a FET Open research and innovation action under the grant agreement No. 964770 (TopoLight), and by the European Union EIC Pathfinder Open project “Polariton Neuromorphic Accelerator” (PolArt, Id: 101130304). Additional support was provided by the ANR Labex GaNext (ANR-11-LABX-0014), the ANR program "Investissements d'Avenir" through the IDEX-ISITE initiative 16-IDEX-0001 (CAP 20-25), the ANR projects MoirePlusPlus (ANR-23-CE09-0033), NEWAVE (ANR-21-CE24-0019), and HAWQ (ANR-25-CE47-7323), and the R\'egion Occitanie. C2N is a member of RENATECH-CNRS, the French national network of large micro-nanofacilities. 
\end{acknowledgments}

\bibliography{biblio}

\section{Supplementary Materials}

\setcounter{equation}{0}
\setcounter{figure}{0}
\setcounter{table}{0}
\setcounter{page}{1}
\setcounter{section}{0}

\renewcommand{\theequation}{S\arabic{equation}}
\renewcommand{\thefigure}{S\arabic{figure}}
\renewcommand{\theHfigure}{S\arabic{figure}}
\renewcommand{\thetable}{S\arabic{table}}
\renewcommand{\thesection}{S\Roman{section}}
\renewcommand{\thepage}{S\arabic{page}}
\renewcommand{\bibnumfmt}[1]{[S#1]} 
\renewcommand{\citenumfont}[1]{S#1}

\subsection{Sample image}

\begin{figure}
    \centering
    \includegraphics[width=\linewidth]{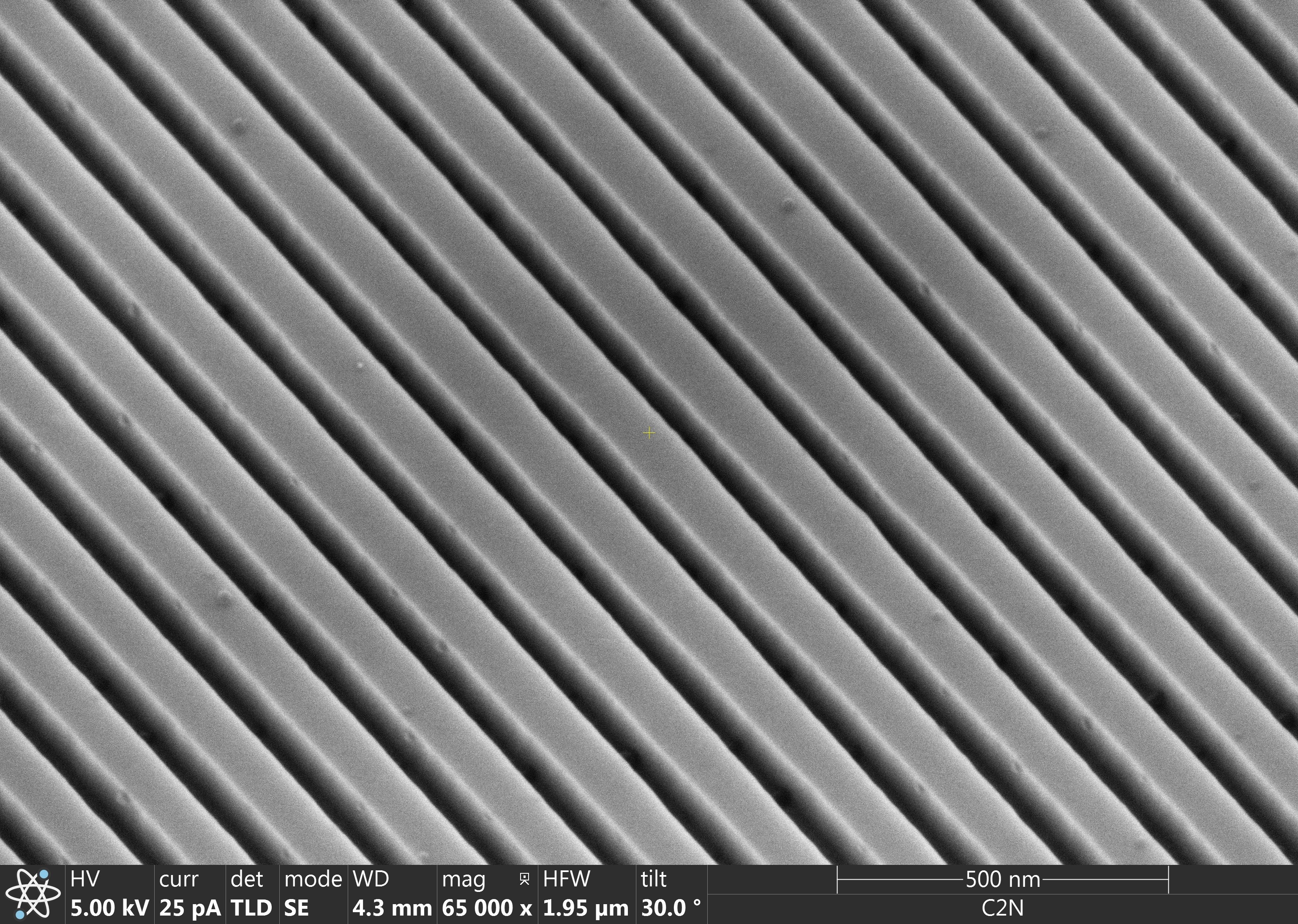}
    \caption{Image of the 1D lattice obtained with an scanning electronic microscope.}
    \label{figMEB}
\end{figure}

Figure~\ref{figMEB} shows scanning electron microscope (SEM) image of the patterned surface of the sample (period $a_0=135 \ nm$, filling factor 60\%). The 1D lattice exhibits a high degree of uniformity over the entire observed area, with a very low density of defects. The tilted view (30$^\circ$) reveals that the etching profile is sufficiently rectangular. This periodic lattice is responsible for the formation of the BIC state. 

\subsection{Polaritonic BIC simulations}

We simulate the photonic structure using COMSOL Multiphysics Wave Optics module. We solve a 2D eigenvalue problem, assuming that the system is homogeneous along the $y$ direction. We use Floquet boundary conditions in the $x$ direction, using $k_x$ as a variable parameter allowing to obtain the dispersion of the eigenvalues. Scattering (reflectionless) boundary conditions are used at the other edges of the system. The simulation allows obtaining the energies, loss rates, and polarizations of the eigenstates. The energies are used to plot the dispersion, the loss rate allows to identify the BIC state (where it is quenched), and the polarization is used in the next section to demonstrate the polarization vortex providing the topological protection to the BIC.

\begin{figure}
    \centering
    \includegraphics[width=0.85\linewidth]{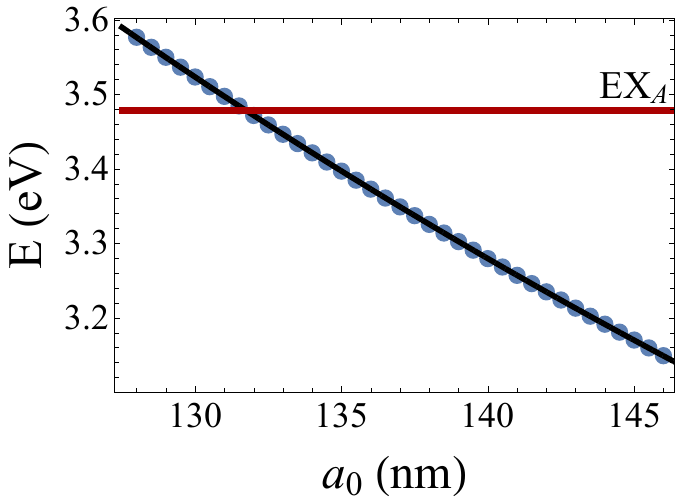}
    \caption{Energies of the photonic BIC state calculated using COMSOL Multiphysics versus period of the grating $a_0$. The red line indicated position of the GaN A-exciton resonance energy. The solid black curve corresponds to the $1/a_0$ fit.}
    \label{figs2}
\end{figure}
The COMSOL simulations have allowed us to perform a scan over parameters of the modulation $a_0$ (period), filling factor, and modulation depth, in order to optimize the energy of the BIC state and put it into resonance with the excitonic reservoir via the LO phonons in order to make the polariton condensation possible and preferential on the BIC state. Figure~\ref{figs2} shows calculated in COMSOL energies of the BIC state $k_x=0$ for the different values of the grating period $a_0$. The black curve corresponds to the hyperbolic fit. Further, the red line indicates the energy of the A-excitons $E_{EX}=3.477$~eV which couples to the considered photonic BIC and lossy modes.

Once the energies of the photonic structure are calculated, we describe the strong light-matter coupling of the obtained photonic modes using a coupled oscillator model, with the Rabi splitting and detuning as two adjustable parameters. Indeed, COMSOL does not allow including a variable refractive index, which is important for a realistic description of the photonic modes in vicinity of the excitonic resonance~\cite{brimont2020strong}. Using a simplified description that does not account for refractive index dispersion generally leads to larger Rabi splitting values, but in general it provides a sufficient precision, especially if the observed phenomena are restricted to a narrow range of energies not too close to the excitonic resonance, as is the case in the present work.

\subsection{BIC polarization vortex}

\begin{figure}
    \centering
    \includegraphics[width=1.0\linewidth]{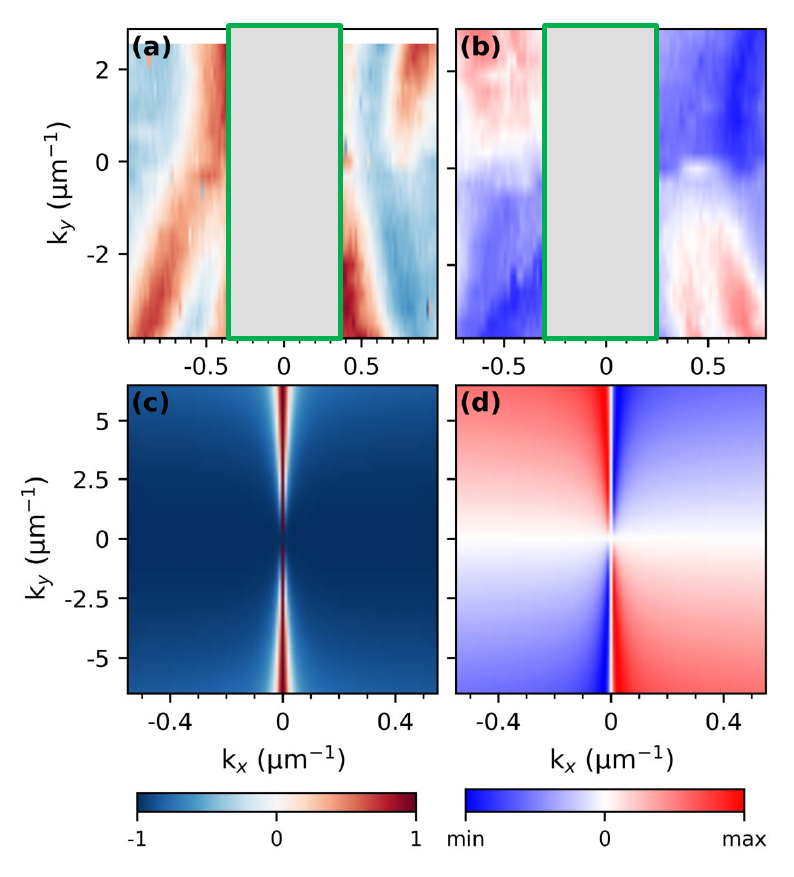}
    \caption{Degrees of linear (a),(c) and diagonal (b),(d) polarizations. Top: extracted from the experiments (the rectangles indicate the regions where the signal is too weak), bottom: extracted from numerical simulations (COMSOL).}
    \label{figs3}
\end{figure}
Figure~\ref{figs3} shows the Stokes vector components $S_1, S_2$ extracted from the experimental PL along the BIC band (a,b) and calculated theoretically with COMSOL Multiphysics (c,d). 
Experimental extraction was performed by fitting the PL spectra at each point in reciprocal space determined by $(k_x,k_y)$ with a Lorentzian function with an offset, and then using the integrals of the extracted Lorentzians to compute the polarization degrees. These projections were used to plot the orientation of the polarization in Fig.~2(b,c) of the main text. The central part of the experimental images is covered by a rectangle to stress that there is no reliable signal coming from the BIC band in this region, so whatever polarization is detected, it is not completely related to that of the BIC band.

\subsection{Localized soliton}

\subsubsection{Soliton size}

\textcolor{black}{In this section, we analyse the profile of the gap soliton wavefunctions in different limits explaining how Eqs.~(1) and (2) of the main text are obtained. We consider the localization induced by a Gaussian reservoir potential, by the self-particle interactions, and the delocalization effect induced by the mixing with the positive mass band. We also discuss how the presence of the BIC affects the emission profile in k and real space with respect to the gap soliton wavefunction.
\subsection{Reservoir-induced localization}
We consider a Gaussian reservoir potential $R(x)=V_0e^{-\frac{x^2}{2\sigma_x^2}}$. The central part of this potential can be approximated by a parabola 
$R(x)\approx V_0-V_0\frac{x^2}{2\sigma_x^2}$.
For negative mass particles, this potential provides parabolic confinement of a harmonic oscillator. The ground state energy is $V_0$ minus the confinement energy $\sqrt{\frac{V_0}{m}}\frac{\hbar}{2\sigma_x}$.
Taking $V_0=$ 3 meV corresponding to the middle of the gap gives 0.4 meV of confinement energy for the large pump spot and 1.2 meV for the small pump spot. In fact, as discussed in the main text, the disorder-induced fluctuations of the reservoir density probably lead to similar potential size in both cases with confinement energy around 1 meV. The corresponding wavefunction for non-interacting particles is a Gaussian with a sigma parameter given by: $\sigma_E=(\frac{\hbar\sigma_x}{\sqrt{mV_0}})^{\frac{1}{2}}$
which gives 5.6 $\mu$m for the large spot and 3.1 $\mu$m for the small spot. The corresponding wavefunction in k-space is also a Gaussian, with a broadening parameter $2\pi/\sigma_E$. Yet, these widths are not directly related to what could be measured in the far-field emission. The emission rate of the BIC band behaves as $k^2$ for the intensity and as $k$ for the amplitude, with a sign change linked to the topological protection of the BIC~\cite{zhen2014topological}. In real space, the Fourier transform of the product between $k$ and a Gaussian gives an intensity $I_R\sim x^2e^{\frac{-x^2}{\sigma_E^2}}$. The RMS width of this function is $\sqrt{3/2}\sigma_E$.}

\textcolor{black}{\subsection{Contribution of the positive mass band}
For the lossy band of energy $E_2$ and mass $m_2$, the profile of the wavefunction at the energy $E$ can be obtained analytically using the WKB (semi-classical) approximation. For this band, the problem is equivalent to considering a solution with an energy below the minimum of the band, and, moreover, in presence of a potential barrier. Such a solution decays exponentially, with a characteristic inverse length given by $\kappa=1/\xi_2=\hbar^{-1}\sqrt{2m_2(E_2+V_0-E)}$.
The expected intensity in real space is $I_2(x)\sim \exp(-2|x|/\xi_2)$.
The intensity profile with the contribution of both bands should read:
$I=\beta I_R+(1-\beta) I_2$,
where $\beta$ is a parameter which should be decreasing with the energy of the gap soliton state.}

\begin{figure}
    \centering
    \includegraphics[width=1.0\linewidth]{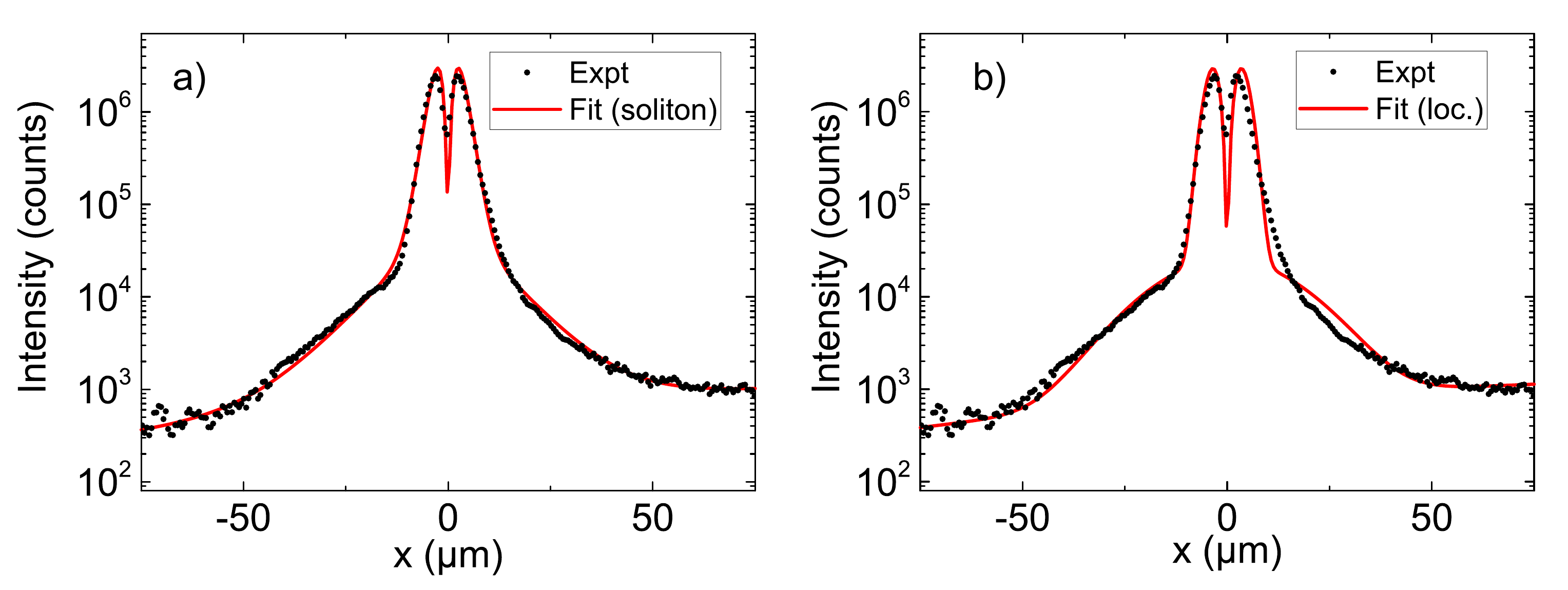}
    \caption{Experimental soliton profile in real space (black dots) and its fit (red solid lines) a)  with a soliton wavefunction, b) with a localized state wavefunction.}
    \label{twofits}
\end{figure}

\textcolor{black}{\subsection{Localisation induced by polariton-polariton interaction}
 A soliton wavefunction driven by interaction reads in real space $\psi_1(x)\sim \sech(x/\xi_1)$ with $\xi_1=\hbar/\sqrt{m\alpha n}$, where $n$ is the particle density at the center of the soliton.
 $\xi_1=3.3 \mu m$, as found in the main text, corresponds to $\alpha n=1.1 meV$, a value which is plausible for the highest pumping power. 
 The Fourier transform of this wavefunction is still a hyperbolic secant: $\psi_1(k)\sim \sech(\frac{\pi\xi_1}{2}k)$. As explained above for the non-interacting case, one needs to multiply this wavefunction by $k$ to get the amplitude of the emitted signal  $k\sech(\frac{\pi\xi_1}{2}k)$. Transforming it back to to real space gives the expected intensity profile $I_S(x)\sim \tanh^2(x/\xi_1)\sech^2(x/\xi_1)$. The corresponding RMSw is around 1.3 $\xi_1$.
 }
 \textcolor{black}{The total intensity including the effect of the second band reads
 $I=\beta I_S+(1-\beta) I_2$
 From these two approaches, we have two approximate possibilities to fit the density profiles corresponding to Eqs.~(1) and (2) of the main text.
 Fig.~\ref{twofits} shows the measured real-space profile of the condensate emission (black dots) for the small spot at $1.3~P_{th}$
 together with two fits (red solid lines): one with the soliton wavefunction described above (panel a) which is also shown in the main text as Fig. 5(d), and the other with the purely localized state wavefunction obtained in a similar way, accounting for the emission properties of the bands (panel b). The first fit gives a better quality: the rms error (square root of the sum of squares due to error divided by the degrees of freedom) is 0.2 for the first fit and 0.32 for the second fit, meaning that the experimental profile is quantitatively closer to the soliton profile than to a purely localized state. The same method is used to estimate the fit quality for Fig. 4 of the main text.}

\subsection{Numerical simulation}
\subsubsection{1D 1 band model}
In order to include both the reservoir and self-interactions self-consistently, we need to perform numerical simulations. We start by presenting a 1D 1-band model which was used to justify the analytical fitting function used in Fig.~4(h). 

The numerical simulations are based on the modified Gross-Pitaevskii equation:\textcolor{black}{
\begin{widetext}
\begin{equation}\label{GPBECBIC}
    i \hbar \frac{\partial}{\partial t} \psi(x,t) = \left ( (i \Lambda - 1) \frac{\hbar^2}{2 m}\frac{\partial^2}{\partial x^2}   + \alpha \left(| \psi (x,t) |^2 +2 n_\mathrm{x}(x)\right) + i\hbar( W_{\mathrm{LO}}n_{\mathrm{x}}(x)e^{-N_{\mathrm{x}}/N_\mathrm{sat}}-  \Gamma_0)  \right ) \psi (x,t)+\chi(x,t).
\end{equation}
\end{widetext}
where $m = -(3.1\pm 0.1)\times 10^{-6} m_0$ is the effective mass of the BIC band, obtained from the experimental polariton dispersion shown in Fig.~1(b) of the main text, $\Lambda = -0.05$ determines the $k^2$ loss profile of the BIC state~\cite{septembre2024soliton,Solnyshkov2014gpe} observed experimentally, $n_\mathrm{x}(x)$ is the reservoir density, which we assume to follow the spatial profile of the pumping laser and to be constant in time. The reservoir depletion is taken into account via the saturated gain term. The filling of the condensate from the excitonic reservoir generated by a non-resonant optical pumping is assumed to be assisted by LO phonons through the term containing the LO phonon scattering matrix element $W_\mathrm{LO}$. The noise term $\chi(x,t)$ allows us to simulate the spontaneous scattering into the polaritonic modes from the reservoir by all mechanisms. }

\begin{figure}
    \centering
    \includegraphics[width=1.0\linewidth]{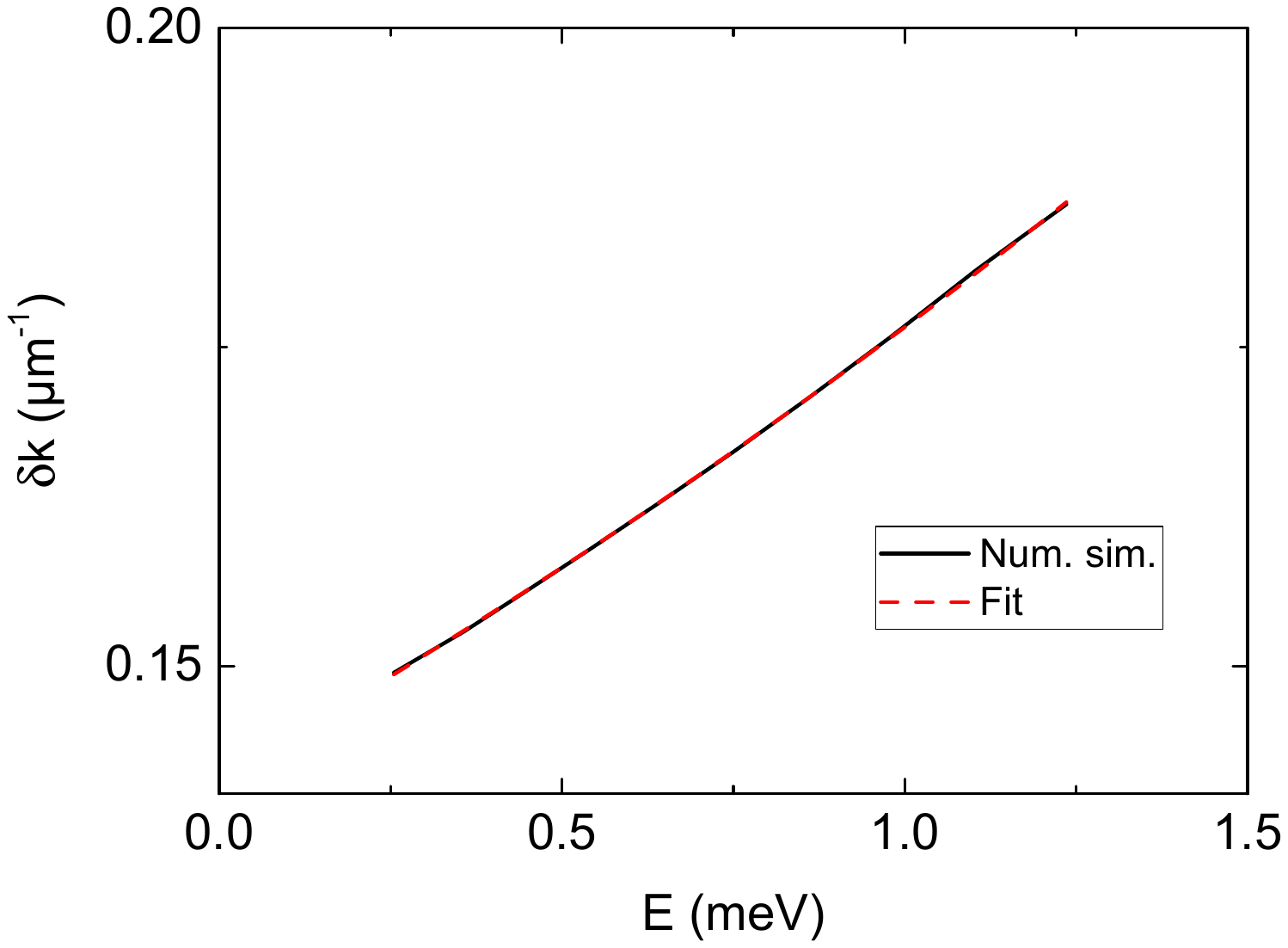}
    \caption{The size of the soliton from simulations (black solid line) together with its analytical fitting function (red dashed line). }
    \label{figSsolth}
\end{figure}

\textcolor{black}{Fig.~4(g) of the main text shows the experimental dependence of the soliton energy on the pumping power. Note that the typical interaction energies observed so far in GaN waveguides are of the order of $\alpha n=0.5$~meV, where $\alpha$ is the effective polariton-polariton interaction constant mostly dominated by the saturation of the exciton oscillator strength and $n$ is their density~\cite{Ciers2020,souissi2024mode}, which is consistent with the blueshift observed in the present experiment in momentum-space measurements of the BIC soliton.
Then we} extracted the soliton size as a function of the interaction energy for different values of the saturation density. The results are plotted in Fig.~\ref{figSsolth} (black solid line).  

The behavior of a localized soliton as a function of its energy can be approximated analytically from the variational approach~\cite{septembre2024soliton} as 
\begin{equation}
\delta k_x \approx \frac{3.6}{\left(\xi(E)-\ell\right)} 
\label{solsize}
\end{equation}
where $\xi(E)=\hbar/\sqrt{mE}$ is the characteristic width that a soliton with the energy $E$ would have, and $\ell$ is a correction corresponding to the localizing potential, which reduces the size of the soliton in real space. The numerical coefficient accounts for the realistic parameters of our system. The value of $\ell$ is obtained by fitting the numerical curve. The result is shown in Fig.~\ref{figSsolth} (red dashed line - analytical fit). The two curves are perfectly overlapping ($R^2=0.9999$), demonstrating the effectiveness of the analytical approximation, at least for the chosen parameters of the reservoir (such as its Gaussian shape). Of course, deviations should be expected at higher energies. \textcolor{black}{The value of $\ell$ used for the fit in the main text is $2.8$~$\mu$m.}
%

\subsubsection{2D 2 bands model }

\textcolor{black}{When the polariton condensation occurs in the second part of the gap which is the case for the experiment done for the small ($9 \mu$m width) pumping spot, the description within the parabolic one-band approximation discussed above becomes incomplete and one needs to take into account a contribution of the upper (lossy) band. This can be naturally done within the basis of two counter-propagating photonic guided modes~\cite{Nigro2023gapstates,ardizzone2022polariton} with linear dispersion, which however implies an equal absolute value of mass for the two resulting bands, whereas in experiment the lossy band is twice heavier.
Moreover, in order to consider the gap soliton dynamics more thoroughly, we extend the 1D model to the actual 2D case, accounting for full hyperbolic dispersion of polaritons for the BIC band, i.e., the transverse $y$ direction, in which the effective mass is positive (contrary to the BIC one -- along the $x$ axis). 
Therefore, the modified Gross-Pitaevskii equations for the forward- and backward-propagating polaritons, $\vec{\psi}=(\psi_+,\psi_-)^T$, under non-resonant pumping (included via the saturated gain term) read as follows:
\begin{widetext}
\begin{equation}\label{2dgpe}
    \imath \hbar \frac{\partial}{\partial t} \vec{\psi}(x,y,t) = \hat{H_1}\vec{\psi}+\left(  \alpha  \vec{\psi}^\dag\vec{\psi} +2\alpha \hat{n}_\mathrm{R}(x,y) + i\hbar W_{\mathrm{LO}}\hat{n}_{\mathrm{R}}(x,y)e^{-N_{\mathrm{x}}/N_\mathrm{sat}}  \right) \vec{\psi}(x,y,t)+\chi(x,y,t).
\end{equation}
\end{widetext}
Here $\hat{H_1}=\hat{H_0}+\hat{H_2}$ with
\begin{equation}\label{h0}
    \hat{H_0}=E_0\pm \hbar v\hat{k}_x+\frac{\hbar^2}{2m_{y}}\hat{k}_y^2,
\end{equation}
describing the energy dispersion of two guided modes. The coupling of these modes is described by $H_2$, which includes the diffractive coupling constant of the grating $U$, and the coupling to the radiative continuum, $\Gamma_r$,
\begin{equation}\label{h2}
    \hat{H}_2=\left(U-\imath \Gamma_r\right)\sigma_x-\imath\Gamma_c\hat{I}-\imath\hat{\Gamma}_Y(k_y)\hat{I}.
\end{equation}
The last two terms in $\hat{H_2}$ represent radiative losses with $\hat{\Gamma}_Y$ related to the $k^2$-losses in the transverse (non-BIC) direction.}

\textcolor{black}{In the absence of an external pump, the eigenenergies of the bands in the small wave vector limit can be written as: 
\begin{equation}\label{eigenv}
    E_{1,2}=E_0+\frac{\hbar^2}{2m_{y}}k_y^2\mp U\left(1\pm \frac{1}{2}\frac{(\hbar vk_x)^2}{U^2+\Gamma_r^2}\right),
\end{equation}
where $1,2$ correspond to the BIC and lossy branches, respectively, $\Gamma_r=\Gamma_c$ in order to have the BIC state at $k=0$~\cite{ardizzone2022polariton}, since
\begin{equation}\label{imagE}
    \mathrm{Im}E_1=-\Gamma_c+\Gamma_r\left(1-\frac{1}{2}\frac{(\hbar vk_x)^2}{U^2+\Gamma_r^2}\right)-\Gamma_Y(k_y).
\end{equation}
}

\textcolor{black}{Numerical integration of the spinor GPE, Eq.~(\ref{2dgpe}), has been performed on a spatial (256x256) grid using multistep Adams-Bashforth method. To increase the performance, we use Graphics Processing Units (GPUs) with PyTorch. The spatial grid cell size is $0.25\mu$m, the time step is $dt=2.67*10^{-4}$ps. 
$\Gamma_r=\hbar/(2\tau)$ with polariton lifetime $\tau=1.6$~ps
,  see the main text. One can note that since we are interested in the evolution of the BIC state, we can ignore the non-parabolicity of the polariton dispersion (at high wave vectors). 
The effective mass obtained from Eq.~(\ref{eigenv}) is $m_{x}=(2.9\pm 1.2)*10^{-6}m_0$
and compares well with the mass obtained fitting the experimental PL (3.1 $10^{-6} m_0$).
The effective mass in the transverse direction $m_y$ is about one order of magnitude bigger as it can be seen from the Fig.~1 of the main text and it is usual case for grating structures. We use $m_y=5.6m_x$ in the numerical simulations. The diffractive coupling constant $U$ is responsible for the gap opening ($2U=6$~meV), see Eq.~(\ref{eigenv}).
}

\subsection{Emission intensity}
\begin{figure}
    \centering
    \includegraphics[width=0.8\linewidth]{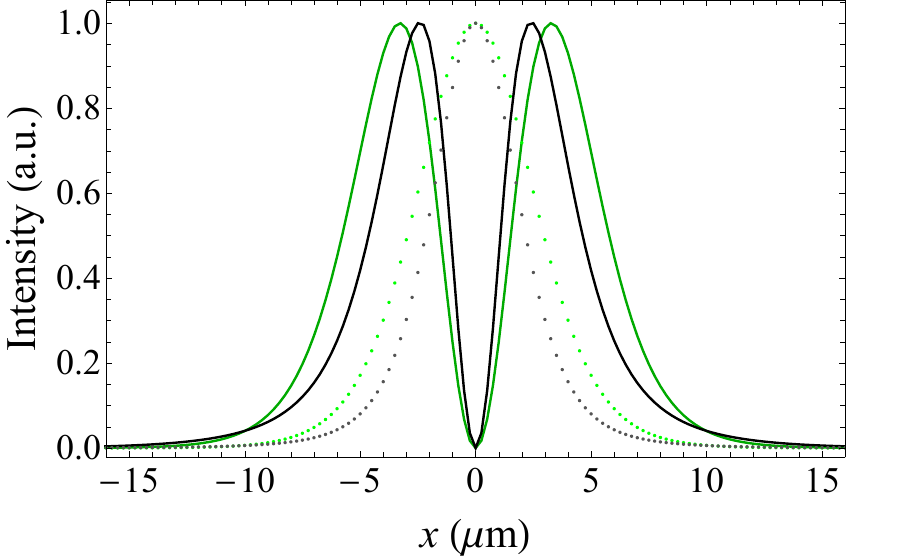}
    \caption{\textcolor{black}{Theoretically calculated intensity profiles for two pumping powers (black curves correspond to the bigger value of the pumping power and subsequently, higher lasing energy) along the BIC direction. Both the internal intensities (dots) and the emission intensities (solid curves) are normalized for the presentation purposes. }}
    \label{figs11}
\end{figure}

\textcolor{black}{Figure~\ref{figs11} shows examples of real space intensity profiles obtained from numerical simulations with Eq.~(\ref{2dgpe}), plotted along the BIC ($x$) direction for two different pumping powers.  The green curves are for the gap soliton with the energy lower than the energy corresponding to the middle of the gap, $E_0$. The black curves correspond to the higher pump and, therefore, to a higher condensate energy (above the middle of the gap). Thus, one can see the change (decrease) in the real space width of the gap soliton with the increase of the energy. Also, we can note a slight change in the shape of the curves (tails), corresponding to the increased contribution of $I_2$. In Fig.~\ref{figs11}, the dotted curves represent the computed probability density (intensity inside the cavity), $I=\vert \psi_+\vert ^2+\vert \psi_-\vert ^2$. The emission intensities (light emitted from the cavity) proportional to the emission rate discussed above for the BIC direction~\cite{sigurdsson2024X,septembre2024soliton} can be found just as $I_{e}=\vert \psi_+ + \psi_-\vert ^2$. Note that all the curves in this figure are normalized by maximal intensity.
}

\begin{figure}
    \centering
    \includegraphics[width=0.8\linewidth]{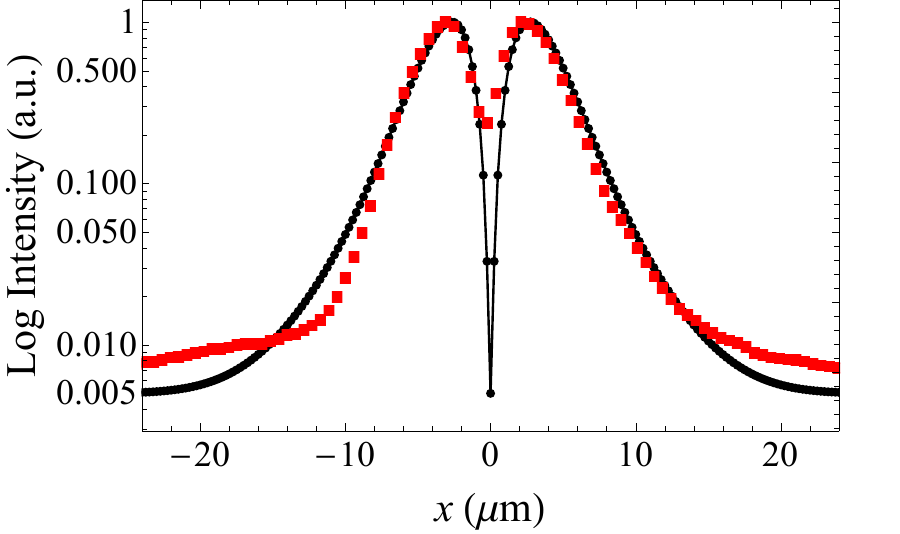}
    \caption{\textcolor{black}{Log-scale intensities profiles (red -- experimental data, see Fig.~\ref{twofits}, black -- theory) along the $x$ direction. The curves correspond to the cuts at $y=0$ for the Fig.~5(b-c) of the main text.
    }}
    \label{figs12}
\end{figure}

\textcolor{black}{Figure~\ref{figs12} shows together two log-scale real space profiles in the BIC direction: experimental data in red (as the black points in Fig.~\ref{twofits}a) and theoretical ones in black (obtained from the simulations of Eq.~(\ref{2dgpe})). 
Here, the theory does not quantitatively reproduce the magnitude of the experimental RMS width. Indeed, our modelling implies equal masses for both the BIC and lossy bands.  As a consequence, the model reproduces well the soliton part of the wavefunction, but not the exponential decay related to the lossy band, for which we use a mass twice lower than the experimental one leading to a faster spatial decay.  Nevertheless, the overall agreement is good, accounting for the background noise in the experimental data which was not subtracted in Fig.~\ref{figs12} (red squares). }

%

\begin{figure}
    \centering
    \includegraphics[width=0.7\linewidth]{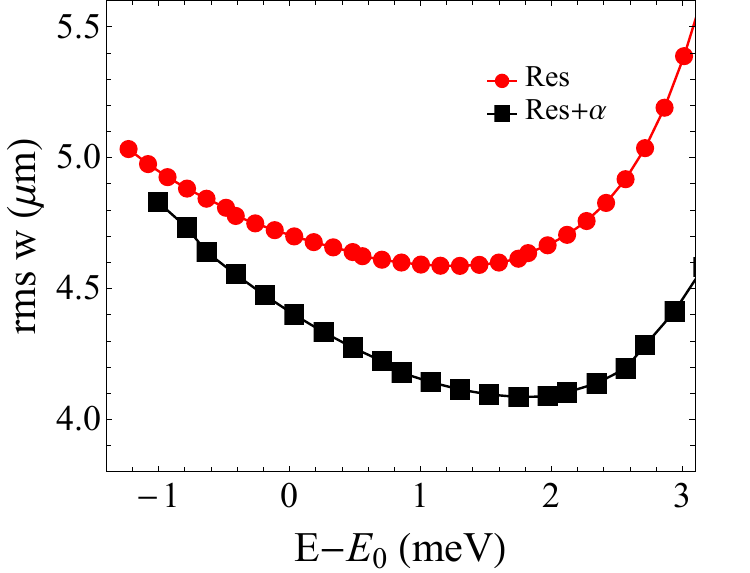}
    \caption{\textcolor{black}{(theory) Roots mean square width (rms w) versus condensate energy (counted from the middle of the gap) in the case of presence (black) and absence of the self-interactions (red). }}
    \label{figs14rmsx}
\end{figure}

\textcolor{black}{As it was discussed above, when the condensation occurs sufficiently close to the lossy band, it starts to contribute to the shape and the width of the gap soliton. Mostly, it can be visible through the exponential tails (still not so visible in the linear scale used in Fig.~\ref{figs11}, since $\beta$ is small). As a result, despite the soliton real space width $\xi_1$ is still decreasing with the energy, the overall rms width can start to grow as it is shown in Fig.~\ref{figs14rmsx}, computed under slightly different conditions in order to show the behavior in a broad energy range. Here red dots correspond to the extracted real space rms width of the gap state induced by the excitonic reservoir only versus the energy of the state, and the black squares mark the width of the gap soliton for the same set of parameters (importantly, the size of the reservoir) but taking in account of the polariton-polariton interactions (characteristic peak interaction energies for the large pumping powers are of the order of $1.5$~meV). Moreover, one can see from the decreased rms width values and the shift of the minimum that localization provided by self-interactions indeed acts in favor of the real space soliton narrowing (k-space soliton broadening). Thus, it becomes essential, since solely excitonic reservoir-induced localization turns out to be not sufficient to get quantitative agreement in rms width with experimental data for the same size of the reservoir.
}

\subsection{Transverse localization}
\begin{figure}
    \centering
    \includegraphics[width=\linewidth]{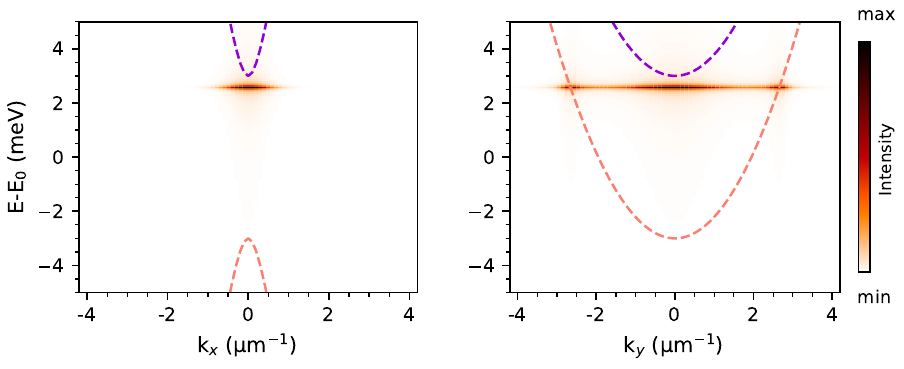}
    \caption{\textcolor{black}{Lasing in the presence of hyperbolicity of the dispersion. Momentum space intensities along the BIC ($x$) and transverse ($y$) directions. The dashed orange and violet curves indicating the BIC and non-BIC (lossy) branches, respectively, added as an guide for the eye. The color scale is adjusted to improve visibility. }}
    \label{figs13}
\end{figure}

The localization in the transverse direction is controlled by the interplay between the mass~\cite{Wouters2008}, the lifetime, and the reservoir potential. The mass in the transverse direction is much larger than that in the longitudinal direction.
\textcolor{black}{Figure~\ref{figs13} shows the momentum space spectral intensities along the $x$ (left panel) and $y$ (right panel) directions, obtained from numerical simulations. The dashed curves represent the eigenenergies calculated for Hamiltonian $H_1$, Eq.~(\ref{h0}-\ref{h2}).
}

\begin{figure}
    \centering
    \includegraphics[width=0.65\linewidth]{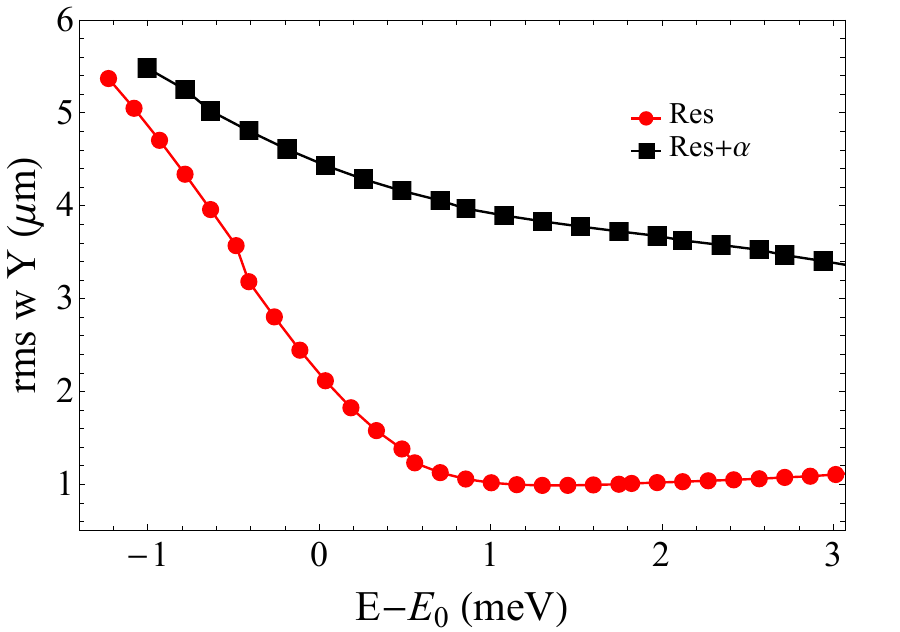}
    \caption{\textcolor{black}{Root mean square width along the transverse direction of the intensity }}
    \label{figs14}
\end{figure}

\textcolor{black}{Figure~\ref{figs14} shows the calculated rms width of the simulated intensity along the $y$ direction in the presence (black squares) and the absence (red dots) of the self-interactions (same conditions as Fig.~\ref{figs14rmsx}). One can see that in contrast to the $x$ direction, repulsive interactions in combination with the positive effective mass for both bands lead to the increase in the width. 
}

\subsection{Soliton excited state at large pump power}

\begin{figure}[h]
    \centering
    \includegraphics[width=0.8\columnwidth]{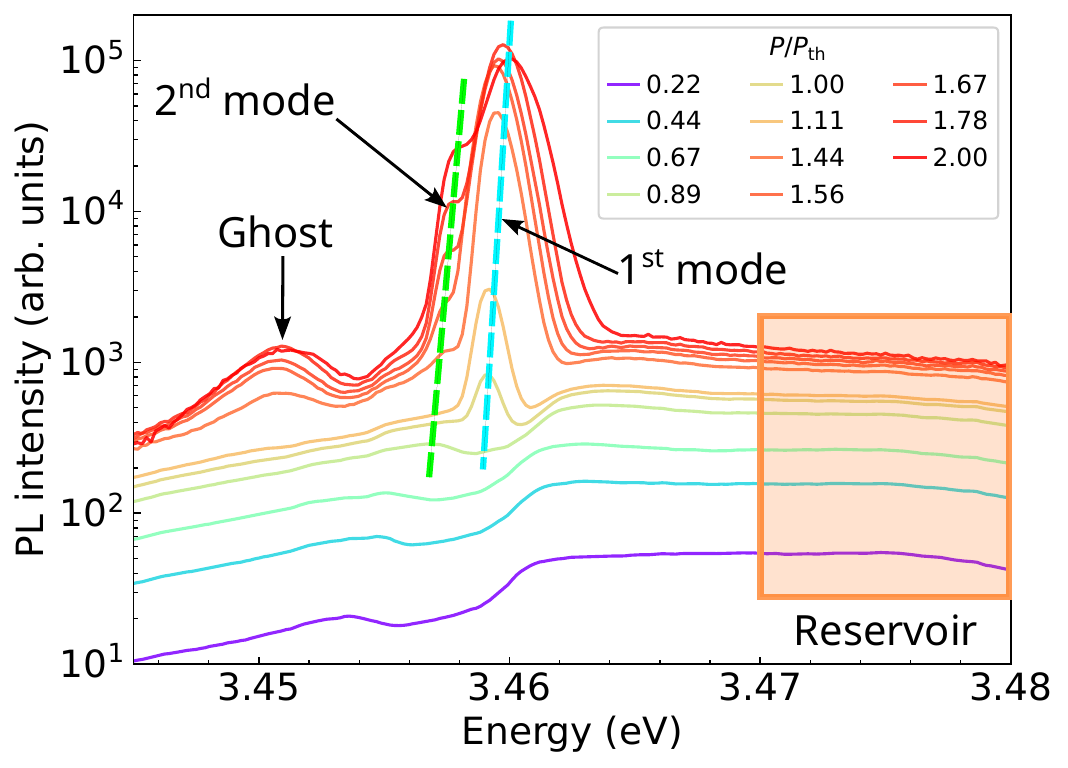}
    \caption{Photoluminescence spectra as a function of excitation power
    $P/P_{th}$. Above threshold, a lasing peak appears at
    $3.460$~eV (fundamental soliton mode), followed beyond
    $1.44\,P_{th}$ by a second mode at $3.458$~eV (p-state).}
    \label{fig:scan_puissance}
\end{figure}

\textcolor{black}{Starting from the soliton image presented in Fig.~5 of the main text for the case of a $9 \mu$m pump spot, the
pump power has been increased up to 2~$P_{th}$. The series of spectra shown in Fig.~\ref{fig:scan_puissance} show the exciton reservoir, the fundamental soliton peak and,
beyond $1.44 P_{th}$, an additional peak at lower energy (3.458~eV), associated to a
second lasing mode. Let us note that the dichroic filter rejecting the pump laser is here tuned
at higher energy than in Fig.~3 of the main text, allowing one to observe the emission from the exciton reservoir.}

\begin{figure}[h]
    \centering
    \includegraphics[width=\columnwidth]{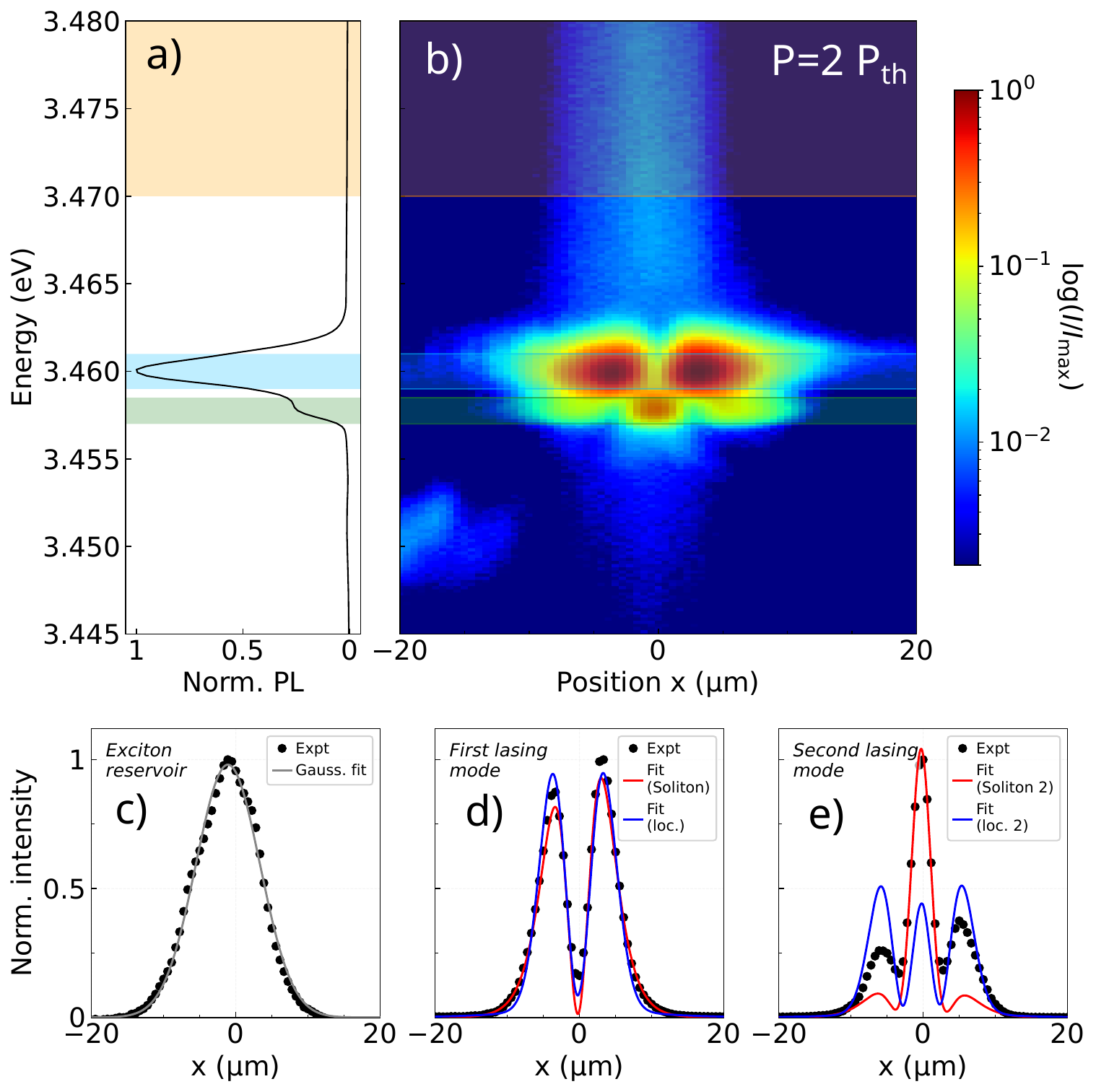}
    \caption{a) Spectrum of the  PL emission at $2P_{th}$, 
    b) Real-space energy-resolved emission map
    (log scale), showing the exciton reservoir (orange zone) the fundamental (cyan zone) and excited (green zone) lasing modes.
    Spatial profiles of the exciton reservoir (Gaussian fit) c), first
    lasing mode d), and second lasing mode e), compared to BIC-filtered soliton fits
    (red) and Hermite-Gauss localized-mode fits (blue).}
    \label{fig:PL_et_coupes}
\end{figure}

\textcolor{black}{The spatial image recorded at the pump power $2P_{th}$ is presented in
figure~\ref{fig:PL_et_coupes}. The profiles extracted from the exciton reservoir, the first and
second lasing modes are shown in panels (c,d,e). To analyse these profiles quantitatively, each
laser mode is fitted with two different functions. The first fit (red line) corresponds
to the BIC-filtered soliton states. The second fit (blue line) corresponds the first excited
state of the harmonic potential, with an Hermite-Gauss profile filtered by the BIC k-dependence
of the losses.}

\textcolor{black}{The fundamental soliton has the spatial profile $I_S(x) + I_{L}(x)$ discussed in the
previous
section. Following the same process, the profile of the first excited state (p-state) is obtained by
transforming the product}
$k \mathrm{sech}\!\left[\frac{k - k_0}{\xi_p}\right] \tanh\!\left[\frac{k - k_0}{\xi_p}\right]$
\textcolor{black}{back to real-space gives the intensity profile $I_{S_2}(x)$ \cite{azzouzi2015} for the first excited
state of the soliton:}
\begin{equation}
    I_{S_2}(x) = \mathrm{sech}^2\!\left(\frac{x - x_0}{\xi_p}\right)
    \left[2\tanh^2\!\left(\frac{x - x_0}{\xi_p}\right) - 1\right]^2
    \label{eq:I2}
\end{equation}
\textcolor{black}{Again, the contribution of the non-BIC band is included with an exponential decay. The fit by
the first excited state of the harmonic potential with Hermite-Gauss profile filtered by the
BIC k-dependance of the losses is also shown in the figure.}

\textcolor{black}{The excited soliton profile reproduces the three-lobe structure of the 2nd lasing mode at
3.458~eV more accurately than the Hermite-Gauss profile, especially the amplitude of the
central lobe, the position of the two nodes and the decay of the lateral tails.}

\textcolor{black}{The observation of the excited state assessed by the fit of its spatial profile is important
because this excited state is at lower energy than the fundamental state. Thus it fully proves
that both lasing modes are hosted by the negative mass BIC branch of the polaritonic device.}

\end{document}